\documentclass[pra,reprint,showpacs,amsmath,amssymb,superscriptaddress]{revtex4-1}
\usepackage{graphicx,bm,color,mathptmx,dcolumn}
\usepackage[colorlinks=true,linkcolor=blue,citecolor=blue,urlcolor =blue]{hyperref}

\newcommand*{\cl}[1]{{\mathcal{#1}}}
\newcommand*{\bb}[1]{{\mathbb{#1}}}

\newcommand{\proj}[2]{| #1 \rangle\!\langle #2 |}
\newcommand*{\tn}[1]{{\textnormal{#1}}}

\newcommand{\T}{\mbox{$\textnormal{Tr}$}}

\newtheorem{Thm}{Theorem}
\newtheorem{Lem}{Lemma}
\newtheorem{Def}{Definition}

\begin{document}

\title{Conditional quantum entropy power inequality for $d$-level quantum systems}

\author{Kabgyun Jeong}
\email{kgjeong6@snu.ac.kr}
\affiliation{Center for Macroscopic Quantum Control, Department of Physics and Astronomy, Seoul National University, Seoul 08826, Korea}
\affiliation{School of Computational Sciences, Korea Institute for Advanced Study, Seoul 02455, Korea}
\affiliation{IMDARC, Department of Mathematical Sciences, Seoul National University, Seoul 08826, Korea}

\author{Soojoon Lee}
\affiliation{Department of Mathematics and Research Institute for Basic Sciences, Kyung Hee University, Seoul 02447, Korea}
\affiliation{School of Computational Sciences, Korea Institute for Advanced Study, Seoul 02455, Korea}

\author{Hyunseok Jeong}
\affiliation{Center for Macroscopic Quantum Control, Department of Physics and Astronomy, Seoul National University, Seoul 08826, Korea}

\begin{abstract}
We propose an extension of the quantum entropy power inequality for finite dimensional quantum systems, and prove a conditional quantum entropy power inequality by using the majorization relation as well as the concavity of entropic functions also given by Audenaert, Datta, and Ozols [J. Math. Phys. \textbf{57}, 052202 (2016)]. Here, we make particular use of the fact that a specific local measurement after a partial swap operation (or partial swap quantum channel) acting only on finite dimensional bipartite subsystems does not affect the majorization relation for the conditional output states when a separable ancillary subsystem is involved. We expect our conditional quantum entropy power inequality to be useful, and applicable in bounding and analyzing several capacity problems for quantum channels.
\end{abstract}

\date{\today}

\maketitle

\section{Introduction}
The channel capacity of a channel (or communication system) in information theory is defined as the maximum rate at which information can be reliably transmitted through the given channel~\cite{S48}. If we choose a communication system such as a quantum mechanical system or quantum channel, which models a quantum state transforming with its ancillary system (or environment), and it is mathematically given by a completely positive, trace-preserving (CPT) map, we can naturally classify quantum, classical and private capacities over the quantum channel according to their respective input information sources ~\cite{NC00,W13}.
In general, determining the channel capacity of a quantum channel is not a simple problem in quantum information theory~\cite{H06}. In particular, it is almost impossible to obtain a channel capacity when quantum entanglement is imposed~\cite{CEM+15}, and most channel capacities are nonadditive~\cite{H09,SY08,LWZG09}.
However, one way to bound the capacity of any channel is to make use of the notion of the entropy power inequality (\textsc{EPI}), originally proposed by Shannon~\cite{S48}. 
In quantum scenarios, \textsc{EPI}s have played a major role in bounding channel capacity for thermal noisy channels (see, for example, Refs.~\cite{KS13,KS13+,BW14}). Furthermore, the concept of \textsc{EPI} is related to a fundamental mathematical isoperimetric inequality in classical as well as quantum regimes~\cite{DPR16}.

First, we briefly review Shannon's statement of the entropy power inequality. The differential entropy for a (continuous) random variable $X$ of values $x\in\bb{R}^d$ with probability density function $p_X$ is defined as~\cite{S48}
\begin{equation}
H(X):
=-\int_{\bb{R}^d} p_X(x)\log p_X(x)\tn{d}^dx,
\end{equation}
which is the relevant information measure for the random variable $X$, and plays a central role in classical information theory. If the random variable $X$ takes a Gaussian distribution $\cl{G}_X$, we can obtain a variance $\tfrac{1}{2\pi e}e^{2H(X)/d}=\nu(\cl{G}_X)$, which is usually called the \emph{entropy power} or energy of the input random variable $X$. For convenience, we omit the factor $\tfrac{1}{2\pi e}$ in the definition of the entropy power. Now, suppose that two independent random variables $X_1$ and $X_2$ on $\bb{R}^d$ are combined via the \emph{scaled addition rule} or the (scaled) convolution operation ($*_t$); then, for a given output signal $X_1*_tX_2$ at the end of the channel, we can find the following classical entropy power inequality (\textsc{cEPI})~\cite{L78,DCT91}:
\begin{equation} \label{eq:cEPI}
\nu(X_1*_tX_2)\ge t\nu(X_1)+(1-t)\nu(X_2),
\end{equation}
where $X_1*_tX_2=\sqrt{t}X_1+\sqrt{1-t}X_2$ is the output signal under the convolution operation with a mixing parameter $t\in[0,1]$. This expression can be restated as the following inequalities:
\begin{align*}
\exp\left(\frac{2H(Y)}{d}\right)&\ge t\exp\left(\frac{2H(X_1)}{d}\right)+(1-t)\exp\left(\frac{2H(X_2)}{d}\right), \\
\textnormal{or}\;\;H(Y)&\ge tH(X_1)+(1-t)H(X_2),
\end{align*}
where $Y:=X_1*_tX_2$.
Details of its proof can be found in several references~ (see \cite{L78,DCT91,B75,BL76,R11,S59,B65}).

Recently, a quantum (Gaussian) version of the entropy power inequality, namely the quantum entropy power inequality (\textsc{qEPI}), has been proved~\cite{KS14,PMG14} and applied to several information-processing tasks~\cite{KS13,GES08,PMLG15}. The \textsc{qEPI} is a quantum analog (but not a direct generalization) of the \textsc{cEPI} equipped with a $\tau$-transmissivity beamsplitter, simply $\tau$-BS of $\tau\in[0,1]$, and whose input sources are $D$-mode bosonic Gaussian quantum states $\rho_{X_\ell}\in\textnormal{Sp}(2D,\bb{R}),~\forall \ell\in\{1,2\}$ on the symplectic space. If we define an entropic function as $\nu_\kappa(\rho_X):=e^{\kappa S(\rho_X)}$, where $S(\varrho)=-\T\varrho\log\varrho$ is the von Neumann entropy of a quantum state $\varrho$, then we have
\begin{equation} \label{eq:QEPI}
\nu_\kappa(\rho_{X_1}\boxplus_\tau\rho_{X_2})\ge\tau\nu_\kappa(\rho_{X_1})+(1-\tau)\nu_\kappa(\rho_{X_2}),
\end{equation}
where $\rho_{X_1}\boxplus_\tau\rho_{X_2}\in\textnormal{Sp}(2D,\bb{R})$ is an output signal of the $\tau$-BS known as the \emph{(Gaussian) quantum addition rule}, and the constant $\kappa=\tfrac{1}{D}$ in the Gaussian case. Generally, the beamsplitter transformation with a parameter $\tau$ can be interpreted as a CPT map $\mathbf{G}_\tau$ over two bosonic modes $\rho_{X_\ell}$ such that
\begin{equation}
\mathbf{G}_\tau:\rho_{X_1}\otimes\rho_{X_2}\mapsto\rho_{X_1}\boxplus_\tau\rho_{X_2}=\T_{X_2}V_\tau(\rho_{X_1}\otimes\rho_{X_2})V_\tau^\dagger,
\end{equation}
where the beamsplitting operation is explicitly given by $V_\tau:=\sqrt{\tau}\openone+\iota\sqrt{1-\tau}\sigma_x$~\cite{ADO16,BCR} including the complex number $\iota=\sqrt{-1}$. We note that $\openone$ is an identity matrix and $\sigma_x$ is the Pauli $x$-matrix, where the $\tau$-BS operation generally interpolates these two operators. Now, if we define $\rho_Y:=\rho_{X_1}\boxplus_\tau\rho_{X_2}$, then we know that \textsc{qEPI}, Eq.~(\ref{eq:QEPI}), has an entropic form of $S(\rho_Y)\ge \tau S(\rho_{X_1})+(1-\tau)S(\rho_{X_2})$ for two independent inputs $\rho_{X_\ell}$ and for the $\tau$-BS. By employing the quantum de Bruijn's inequality and the entropy-scaling property known as `Gaussification,' we can obtain the entropic inequality~\cite{KS14}---the entropy of a channel's mixed output is always increased.

A \textsc{qEPI} for $d$-dimensional quantum states (qudits) has also been proposed~\cite{ADO16}, and is given by the form of Eq.~\eqref{eq:QEPI}, but it is generally true when the constant $\kappa$ is restricted to $0\le\kappa\le\tfrac{1}{(\log d)^2}$ where $d\simeq2D$. In the proof, the symmetric property and the concavity of the entropic function $\nu_\kappa(\rho)$ in the region of $\kappa$ via the majorization relation on a quantum state $\rho$ was used. Furthermore, it is important to note that independent input quantum states for the quantum channel are represented by $\rho_{X_\ell}\in\bm{D}(\bb{C}^d)$ with $\ell\in\{1,2\}$, where $\bm{D}(\bb{C}^d):=\{\rho\in\bm{B}(\bb{C}^d):\T\rho=1,\rho=\rho^\dagger\ge0\}$ is a class of density matrices on a bounded linear operator $\bm{B}(\bb{C}^d)$ (over the $d$-dimensional Hilbert space), and those mixing operations with the parameter $\tau$ are given by a \emph{partial swap} as follows.
We now review the partial swap operation ($p$-\textsc{Swap}) denoted by $\boxplus_\tau$, which is also known as the \emph{qudit addition rule}~\cite{ADO16}. For any $\tau\in[0,1]$ and any density matrices $\rho_{X_\ell}\in\bm{D}(\bb{C}^d)$, we can find an output of the quantum channel via the $p$-\textsc{Swap} as
\begin{align} 
\rho_{X_1}\boxplus_\tau\rho_{X_2}&=\mathbf{N}_\tau(\rho_{X_1}\otimes\rho_{X_2}) \nonumber\\
&=\T_{X_2}\left[U_\tau(\rho_{X_1}\otimes\rho_{X_2})U_\tau^\dagger\right] \nonumber\\
&=\tau\rho_{X_1}+(1-\tau)\rho_{X_2}-\iota\sqrt{\tau(1-\tau)}[\rho_{X_1},\rho_{X_2}], \label{eq:partialSW}
\end{align}
where $[A,B]=AB-BA$ is the commutator, the resulting state is also a $d$-level quantum state, and $U_\tau:=\sqrt{\tau}\openone+\iota\sqrt{1-\tau}W$, where $W$ is the swap operator such that $W\rho_{AB}W^\dagger=\rho_{BA}$ on two $d$-level quantum systems. We call the map $\mathbf{N}_\tau(\cdot)$ the \emph{partial swap channel} on $d$-level quantum systems.

In this study, we prove a conditional version of the \textsc{qEPI} (\textsc{CqEPI}) for arbitrary $d$-level quantum states in Sec.~\ref{main} through a conditional majorization relation (see Sec.~\ref{majorization}). We discuss our results and outline our future plans in Sec.~\ref{discussion}.

\section{Conditional eigenvalues and majorization relation for quantum states} \label{majorization}
It was conjectured that, for any quantum states $\rho_{X_1X_2E}$ and a mixing parameter $\tau\in[0,1]$,
\begin{equation} \label{open1}
S(\rho_{X_1}\boxplus_\tau\rho_{X_2}|\rho_E)\ge\tau S(\rho_{X_1}|\rho_E)+(1-\tau)S(\rho_{X_2}|\rho_E),
\end{equation}
where the beamsplitter operation with $\tau$ acts on any two quantum systems~\cite{K15}. However, for any Gaussian product states---especially having the form $\rho_{X_1E_1}\otimes\rho_{X_2E_2}$, Koenig proved that $S(\rho_Y|\rho_{E})\ge\tau S(\rho_{X_1}|\rho_{E_1})+(1-\tau)S(\rho_{X_2}|\rho_{E_2})$, where $\rho_Y=\rho_{X_1}\boxplus_\tau\rho_{X_2}$ and $\rho_{E}=\rho_{E_1}\otimes\rho_{E_2}$ is the (separable) ancillary system. Koenig referred to this inequality as the \emph{conditional} quantum \textsc{EPI} or \textsc{CqEPI} in the Gaussian regime. In his proof, Koenig exploits the quantum version of the ``scaling property for the conditional entropy'' (Lemma 6.2 in Ref.~\cite{K15}) and the ``conditional de Bruijn identity'' (Theorem 7.3 also in Ref.~\cite{K15}) in the Gaussian regime. Recently, a similar result for the Gaussian \textsc{CqEPI} is introduced by de Palma and Trevisan~\cite{PT17}. In their papers, they have used quantum conditional entropy notation, $S(\rho_A|\rho_B)=S(A|B)_{\rho_{AB}}:=S(AB)_{\rho_{AB}}-S(B)_{\rho_B}$, which means the von Neumann entropy of system $A$ when system $B$ is conditioned. However, in this paper, we use a different notation of a set of conditional eigenvalues such as $\lambda(\rho_A|_B)$, given by any quantum measurement performed on the subsystem $B$, so as to show another version of the \textsc{CqEPI} based on local measurements, which is not the same as the \textsc{CqEPI} with respect to the quantum conditional entropy. 
Our approach is related to the quantum discord, which represents another type of quantum correlation different from entanglement~\cite{OZ01,HV01,DSC08,AD10,MBC+12,GTA13}. 

The Gaussian \textsc{CqEPI} comes from the fact that, if \emph{any} quantum state $\rho_{X_1X_2E}$ has a conditionally independent form, i.e., $\rho_{X_1X_2E}=\rho_{X_1E_1}\otimes\rho_{X_2E_2}$, then it can be decomposed as a direct sum of tensor products~\cite{K15,HJPW04} such that
\begin{equation}
\rho_{X_1X_2E}=\bigoplus_{j}p_j\rho_{X_1E_1^j}\otimes\rho_{X_2E_2^j},
\end{equation}
and the von Neumann entropy of state $\rho_{X_1X_2E}$ satisfies $S(\bigoplus_jp_j\rho_j)=\sum_jp_jS(\rho_j)+H(\{p_j\}_j)$, where $H(\cdot)$ is the Shannon entropy~\cite{SSA}. Instead of Gaussian product states, we give a similar proof of the \textsc{qEPI} for any $d$-level product states $\rho_{X_1E_1}\otimes\rho_{X_2E_2}$ conditioned through a quantum measurement on the environments $E_1$ and $E_2$ respectively. For $d$-level \textsc{CqEPI} cases, we use the majorization relation for eigenvalues of $\rho_{X_\ell}|_{E_\ell}\in\bm{D}(\bb{C}^{d})$ for all $\ell=1,2$, instead of the quantum conditional entropy. 

Before the main proof, we briefly review the majorization condition for quantum states. Let us denote $\mathbf{m}=(m_1,m_2,\ldots,m_d)$ and $\mathbf{n}=(n_1,n_2,\ldots,n_d)\in\bb{R}^d$ with its components arranged in decreasing order of $m_1^\downarrow\ge m_2^\downarrow\ge\cdots\ge m_d^\downarrow$ and $n_1^\downarrow\ge n_2^\downarrow\ge\cdots\ge n_d^\downarrow$. Then, for any $\mathbf{m}$ and $\mathbf{n}\in\bb{R}^d$, $\mathbf{m}$ is considered to be \emph{majorized} by $\mathbf{n}$ and we write $\mathbf{m}\prec\mathbf{n}$ if, $\forall k=\{1,\ldots,d\}$, $\sum_{j=1}^km_j^\downarrow\le\sum_{j=1}^kn_j^\downarrow$ with equality at $k=d$. In addition, a function $f:\bb{R}^d\to\bb{R}$ is called \emph{Schur concave}, if $f(\mathbf{m})\ge f(\mathbf{n})$ whenever $\mathbf{m}\prec\mathbf{n}$~\cite{B97}. The majorization technique explained above is also obvious in the density operator formalism of the quantum regime~\cite{NC00}.

By using the definition of the majorization condition above, and the partial swap channel in Eq.~\eqref{eq:partialSW}, it was proved in Refs.~\cite{ADO16,CLL16} that, for any quantum states $\rho_{X_1},\rho_{X_2}\in\bm{D}(\bb{C}^d)$, we can obtain 
\begin{equation} \label{eq:major1}
\lambda(\rho_{X_1}\boxplus_\tau\rho_{X_2})\prec\tau\lambda(\rho_{X_1})+(1-\tau)\lambda(\rho_{X_2}),
\end{equation}
where $\lambda(\rho)$ denotes a set of the eigenvalues for a quantum state $\rho$, and $\boxplus_\tau$ the $p$-\textsc{Swap} operation with a mixing parameter $\tau\in[0,1]$. This point is crucial. Our main goal in this study is to extend Eq.~\eqref{eq:major1} to the (measurement-based) conditional version for $d$-level quantum states.

\section{\textsc{CqEPI}: Main results} \label{main}
We now suggest that the $p$-\textsc{Swap} and its identity (Theorem 1.1 in Ref.~\cite{CLL16}) can be extended to a conditional version of the entropy power inequality. Here, we make use of the fact that any local measurements (LMs) via the partial swap operation do not change the majorization condition when the separable environments $E_1$ and $E_2$ are measured locally (see Fig.~\ref{fig:setup} and Lemma~\ref{Lem:main} below). Note that, if $\rho_E\neq\rho_{E_1}\otimes\rho_{E_2}$, the \textsc{CqEPI} is still open as in Eq.~(\ref{open1}). 

\begin{figure}
\includegraphics[width=\columnwidth]{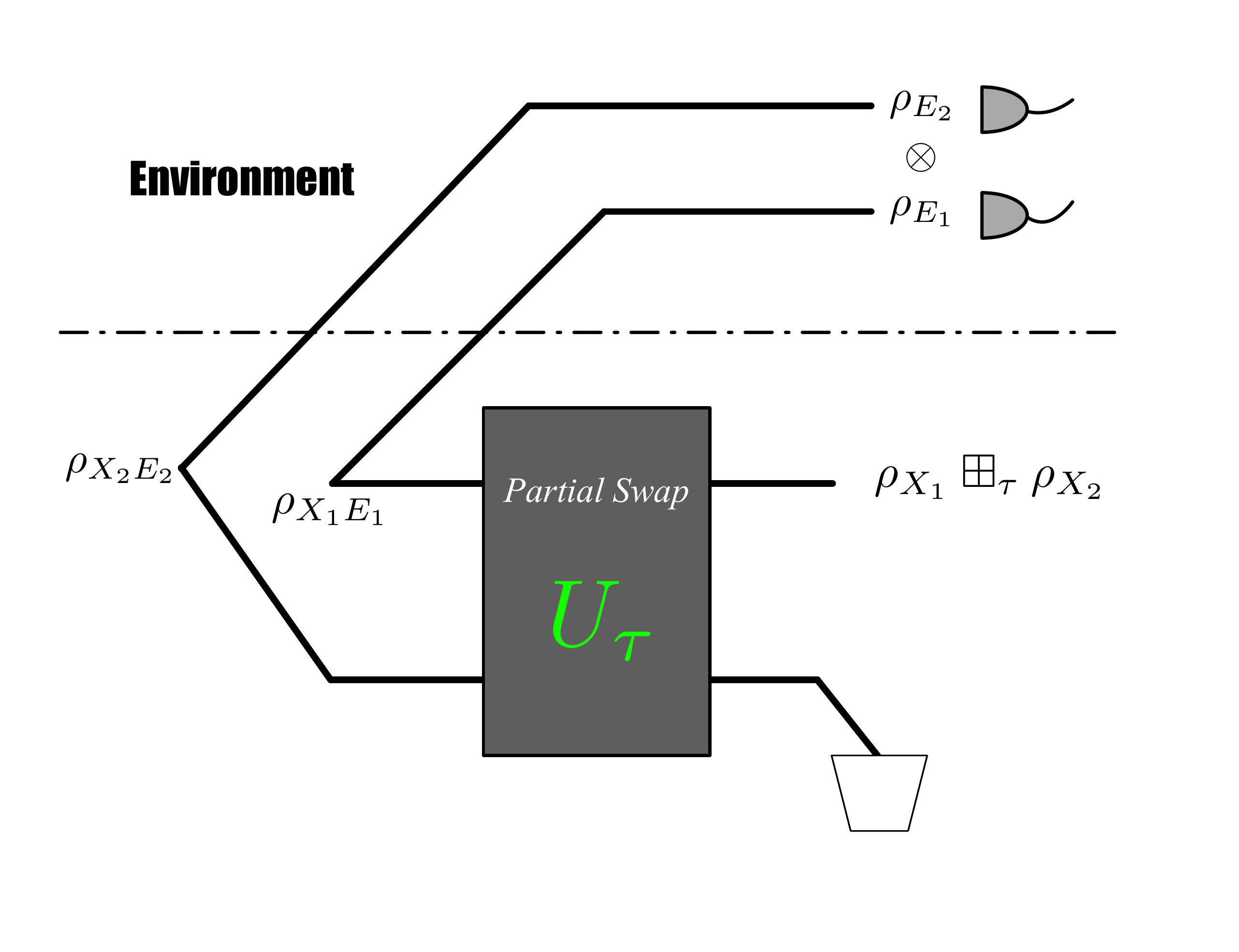}
\caption{ The setting for \textsc{CqEPI} on $d$-level quantum states (qudits). For any product input states in the form of $\rho_{X_1E_1}\otimes\rho_{X_2E_2}$, the diagram represents a quantum channel generating output of $\rho_{X_1}\boxplus_\tau\rho_{X_2}$ for the quantum states. The unitary operation $U_\tau$ corresponds to the $p$-\textsc{Swap} across the two independent inputs $\rho_{X_1}$ and $\rho_{X_2}$ conditioned via quantum measurements on the (separable) environmental subsystems $\rho_{E_1}$ and $\rho_{E_2}$ respectively.}
\label{fig:setup}
\end{figure}

First, we briefly review the output states of the quantum channel through the partial swap operation. Let $\rho_{X_1X_2E_1E_2}:=\rho_{X_1E_1}\otimes\rho_{X_2E_2}$ be the total quantum state. Then we have
\begin{align}
\rho_{YE_1E_2}&=(\mathbf{N}_{\tau}\otimes\openone_{E_1E_2})(\rho_{X_1X_2E_1E_2}) \nonumber\\
&=\T_{X_2}(U_\tau\otimes\openone_{E_1E_2})(\rho_{X_1E_1}\otimes\rho_{X_2E_2})(U_\tau^\dagger\otimes\openone_{E_1E_2}^\dagger), \label{eq:bilocal}
\end{align}
and also remember $\rho_Y=\rho_{X_1}\boxplus_\tau\rho_{X_2}=\mathbf{N}_\tau(\rho_{X_1}\otimes\rho_{X_2})=\tau\rho_{X_1}+(1-\tau)\rho_{X_2}-\iota\sqrt{\tau(1-\tau)}[\rho_{X_1},\rho_{X_2}]$. We now introduce a new set of eigenvalues of $\rho_Y$ induced by $\rho_{YE_1E_2}$ after local measurements on the separable environment $\rho_{E_1}\otimes\rho_{E_2}$, and we will use the notation such as $\lambda(\rho_Y|_{E_1E_2})$. Notice that the notation $\rho_X|_{E}$ does not mean the conditional quantum state introduced in Ref.~\cite{LS13}, but (as mentioned above) it is just a quantum state $\rho_X$ after a local measurement performed on the subsystem $E$ for $\rho_{XE}$. For example, if we choose a set of local measurement described by $\{M_j\}_{E}$ on the subsystem $\rho_{E}$ ($1\le \forall j\le d_E$), then we define
\begin{align}
\rho_{X}|_{E(j)}=\frac{1}{p_j}\T_{E}\left((\openone_X\otimes M_j)\rho_{XE}(\openone_X\otimes M_j^{\dag})\right),
\end{align}
where $p_j=\T(M_j^{\dag}M_j\rho_{E})$ is the normalization factor. Thus, we can naturally define the set of conditional eigenvalues after a specific local measurement on $E$ as follow: ($\forall\rho_{XE}$)
\begin{equation}
\lambda(\rho_{X}|_{E(j)}):=\lambda(\T_{E}[(\openone_X\otimes M_j)\rho_{XE}(\openone_X\otimes M_j^{\dag})]/p_j). \label{eq:majorcon} 
\end{equation}

As a subsidiary example, let us consider $\rho_{YE_1E_2}=(\mathbf{N}_{\tau}\otimes\openone_{E_1E_2})(\rho_{X_1E_1}\otimes\rho_{X_2E_2})$ and a situation in which local projective measurements are involved. Let $\{\proj{\psi_j}{\psi_j}_{E_1}\}_{j=1}^{d_{E_1}}$ and $\{\proj{\phi_k}{\phi_k}_{E_2}\}_{k=1}^{d_{E_2}}$ be the local measurements on the environmental subsystems $\rho_{E_1}$ and $\rho_{E_2}$ respectively. Finally, to find the conditional eigenvalues, we define the final states (conditional outputs) after local measurements on the subsystems $E_1$ and $E_2$ as $\sigma_{X_1}^{(j)}=\frac{1}{q_j^{(1)}}{}_{E_1}\!\langle\psi_j|\rho_{X_1E_1}|\psi_j\rangle_{E_1}$ and $\sigma_{X_2}^{(k)}=\frac{1}{q_k^{(2)}}{}_{E_2}\!\langle\phi_k|\rho_{X_2E_2}|\phi_k\rangle_{E_2}$ where $q_j^{(1)}={}_{E_1}\!\langle\psi_j|\rho_{E_1}|\psi_j\rangle_{E_1}$ and $q_k^{(2)}={}_{E_2}\!\langle\phi_k|\rho_{E_2}|\phi_k\rangle_{E_2}$. Then
\begin{widetext}
\begin{align*}
\sigma_Y^{(j,k)}&:=\frac{1}{p_{j,k}^{(1,2)}}{}_{E_1E_2}\langle\psi_j,\phi_k|\rho_{YE_1E_2}|\psi_j,\phi_k\rangle_{E_1E_2} \\
&=\left(\mathbf{N}_{\tau}\otimes\openone_{E_1E_2}\right)\left(\frac{1}{q_j^{(1)}}{}_{E_1}\!\langle\psi_j|\rho_{X_1E_1}|\psi_j\rangle_{E_1}\otimes\frac{1}{q_k^{(2)}}{}_{E_2}\!\langle\phi_k|\rho_{X_2E_2}|\phi_k\rangle_{E_2}\right) \\
&=\mathbf{N}_{\tau}\left(\sigma_{X_1}^{(j)}\otimes\sigma_{X_2}^{(k)}\right) \\
&=\sigma_{X_1}^{(j)}\boxplus_\tau\sigma_{X_2}^{(k)}.
\end{align*}
\end{widetext}
Note that $p_{j,k}^{(1,2)}:={}_{E_1E_2}\!\langle\psi_j,\phi_k|\rho_{E_1E_2}|\psi_j,\phi_k\rangle_{E_1E_2}={}_{E_1}\!\langle\psi_j|\rho_{E_1}|\psi_j\rangle_{E_1}\cdot{}_{E_2}\!\langle\phi_k|\rho_{E_2}|\phi_k\rangle_{E_2}=q_j^{(1)}\cdot q_k^{(2)}$, since $\rho_{E_1E_2}=\rho_{E_1}\otimes\rho_{E_2}$ is separable.

By using Theorem 1.1 in Ref.~\cite{CLL16}, we can naturally obtain that
\begin{equation}
\lambda(\sigma_{X_1}^{(j)}\boxplus_\tau\sigma_{X_2}^{(k)})\prec\tau\lambda(\sigma_{X_1}^{(j)})+(1-\tau)\lambda(\sigma_{X_2}^{(k)}).
\end{equation}
This relation directly implies that specific local measurements after the $p$-\textsc{Swap} operation do not affect the majorization relation for the conditional output states. Without loss of generality, we can generalize the (local) projective measurement to a (local) general measurement formalism. For the main proof, we need the following definition, which is a natural extension of Eq.~(\ref{eq:partialSW}) (see also Fig.~\ref{fig:setup}).

\begin{Def}[Output state of $p$-\textsc{Swap} operation] \label{Def:sub}
For any quantum states in the form $\rho_{X_1X_2E_1E_2}:=\rho_{X_1E_1}\otimes\rho_{X_2E_2}$, the output state through the partial swap operation with $\tau\in[0,1]$ on subsystems $X_1$ and $X_2$ is given by
\begin{equation} \label{eq:monotone}
\rho_{YE_1E_2}=\tau\rho_{X_1E_1}+(1-\tau)\rho_{X_2E_2}-\iota\sqrt{\tau(1-\tau)}[\rho_{X_1E_1},\rho_{X_2E_2}].
\end{equation}
\end{Def}

By using Definition~\ref{Def:sub} and Eq.~(\ref{eq:majorcon}), we can derive the following crucial lemma, namely the `conditional majorization relation' for our product $d$-level quantum states. First, we define $\rho_{X_1}|_{E_1(j)}:=\frac{1}{q_j^{(1)}}\T_{E_1}(M_j^{(1)}\rho_{X_1E_1}M_j^{\dag(1)})$ and $\rho_{X_2}|_{E_2(k)}:=\frac{1}{q_k^{(2)}}\T_{E_2}(M_k^{(2)}\rho_{X_2E_2}M_k^{\dag(2)})$, i.e., the outcome states after local measurements given by $\{M_j^{(1)}\}_{E_1}$ and $\{M_k^{(2)}\}_{E_2}$, where $q_j^{(1)}=\T(M_j^{\dag(1)}M_j^{(1)}\rho_{E_1})$ and $q_k^{(2)}=\T(M_k^{\dag(2)}M_k^{(2)}\rho_{E_2})$ on the environmental subsystems $\rho_{E_1}$ and $\rho_{E_2}$ respectively. Note that, for any $j$, the measurement elements satisfy $\sum_{j=1}^dM_j^\dag M_j=\openone$.

\begin{Lem}[Conditional majorization relation] \label{Lem:main}
For any pair of density matrices $\rho_{X_1E_1},\rho_{X_2E_2}\in\bm{D}(\bb{C}^{d\times d_{E_\ell}})$, any $\tau\in[0,1]$ and for all $j,k$, if we take local measurements as $\{M_j^{(1)}\}_{E_1}$ and $\{M_k^{(2)}\}_{E_2}$ on the subsystems $\rho_{E_1}$ and $\rho_{E_2}$ respectively, then we have
\begin{equation} \label{eq:minor}
\rho_{Y}|_{E_1(j)E_2(k)}=\rho_{X_1}|_{E_1(j)}\boxplus_\tau\rho_{X_2}|_{E_2(k)}.
\end{equation}
This fact directly implies that, for each measurement outcome $j$ and $k$,
\begin{equation} \label{eq:major}
\lambda(\rho_Y|_{E_1(j)E_2(k)})\prec\tau\lambda(\rho_{X_1}|_{E_1(j)})+(1-\tau)\lambda(\rho_{X_2}|_{E_2(k)}).
\end{equation} 
Here, the environmental subsystem is given by $\rho_{E_1E_2}=\rho_{E_1}\otimes\rho_{E_2}$, i.e., the separable state.
\end{Lem}

\emph{Proof}. It is sufficient to prove that, for each $j$ and $k$, $\rho_{Y}|_{E_1(j)E_2(k)}=\rho_{X_1}|_{E_1(j)}\boxplus_\tau\rho_{X_2}|_{E_2(k)}$. That is,
\begin{widetext}
\begin{align*}
\rho_{Y}|_{E_1(j)E_2(k)}
&:=\frac{1}{p_{j,k}^{(1,2)}}\T_{E_1E_2}\left((M_j^{(1)}\otimes M_k^{(2)})\rho_{YE_1E_2}(M_j^{\dag(1)}\otimes M_k^{\dag(2)})\right) \\
&=\left(\mathbf{N}_{\tau}\otimes\openone_{E_1E_2}\right)\left(\frac{1}{q_j^{(1)}}\T_{E_1}(M_j^{(1)}\rho_{X_1E_1}M_j^{\dag(1)})\otimes\frac{1}{q_k^{(2)}}\T_{E_2}(M_k^{(2)}\rho_{X_2E_2}M_k^{\dag(2)})\right) \\
&=\mathbf{N}_{\tau}\left(\rho_{X_1}|_{E_1(j)}\otimes\rho_{X_2}|_{E_2(k)}\right) \\
&=\rho_{X_1}|_{E_1(j)}\boxplus_\tau\rho_{X_2}|_{E_2(k)},
\end{align*} 
\end{widetext}
where we again use the fact that the probability $p_{j,k}^{(1,2)}:=\T_{E_1E_2}\left((M_j^{\dag(1)}M_j^{(1)}\otimes M_k^{\dag(2)}M_k^{(2)})\rho_{E_1E_2}\right)=\T(M_j^{\dag(1)}M_j^{(1)}\rho_{E_1})\cdot\T(M_k^{\dag(2)}M_k^{(2)}\rho_{E_2})=q_j^{(1)}\cdot q_k^{(2)}$ for the (separable) environmental system $\rho_E=\rho_{E_1}\otimes\rho_{E_2}$. Second parts (i.e., Eq.~(\ref{eq:major})) are directly given by Theorem 11 in Ref.~\cite{ADO16} or Theorem 1.1 in Ref.~\cite{CLL16}. This completes the proof. $\blacksquare$

In the proof of Lemma~\ref{Lem:main}, for any Schur concave function $f$, 
we can define its function values as 
\begin{align*}
f(\rho_Y|_{E_1E_2})&=
\min_{\{M_j^{(1)}\},\{M_k^{(2)}\}}\sum_{j,k}q_j^{(1)}q_k^{(2)}f(\rho_Y|_{E_1(j)E_2(k)}),\\
f(\rho_{X_1}|_{E_1})&=
\min_{\{M_j^{(1)}\}}\sum_{j}q_j^{(1)}f(\rho_{X_1}|_{E_1(j)}),\;\;\textnormal{and}\\
f(\rho_{X_2}|_{E_2})&=
\min_{\{M_k^{(2)}\}}\sum_{k}q_k^{(2)}f(\rho_{X_2}|_{E_2(k)}).
\end{align*}
Then by exploiting Lemma~\ref{Lem:main}, we can prove the following theorem, which is our main result.

\begin{Thm}[Conditional qudit EPI (\textsc{CqEPI})] \label{Thm:quditCEPI}
Let $\rho_{X_1E_1}$ and $\rho_{X_2E_2}$ be any discrete $d\times d_{E_\ell}$-level quantum states with a separable environment $\rho_{E_1}\otimes\rho_{E_2}$ and $\ell\in\{1,2\}$. For any concave and symmetric function $\nu_\kappa$ with a range of $0\le\kappa\le\tfrac{1}{(\log d)^2}$, and for any $\tau\in[0,1]$, we have
\begin{equation}
\nu_\kappa(\rho_Y|_{E_1E_2})\ge\tau \nu_\kappa(\rho_{X_1}|_{E_1})+(1-\tau)\nu_\kappa(\rho_{X_2}|_{E_2}).
\end{equation}
\end{Thm}

\emph{Proof}.
For each measurement outcome $j$ and $k$, let $\rho_{X_1}'|_{E_1},\rho_{X_2}'|_{E_2}\in\bm{D}(\bb{C}^d)$ be diagonal states whose entries are the eigenvalues of $\rho_{X_1}|_{E_1}$ and $\rho_{X_2}|_{E_2}$, respectively, arranged in decreasing order. Since $\lambda(\rho_{X_1}'|_{E_1})=\lambda(\rho_{X_1}|_{E_1})$ and $\lambda(\rho_{X_2}'|_{E_2})=\lambda(\rho_{X_2}|_{E_2})$, we then have, from Eq.~\eqref{eq:major},
\begin{align*}
\lambda(\rho_Y|_{E_1E_2})
&\prec\tau\lambda(\rho'_{X_1}|_{E_1})+(1-\tau)\lambda(\rho'_{X_2}|_{E_2}) \\
&=\lambda\left(\tau\rho'_{X_1}|_{E_1}+(1-\tau)\rho'_{X_2}|_{E_2}\right).
\end{align*}
For any entropic function $\nu_\kappa(\cdot)$ that is symmetric and concave in terms of eigenvalues of density matrices, we have
\begin{align*}
\nu_\kappa(\rho_{X_1}|_{E_1}\boxplus_\tau\rho_{X_2}|_{E_2})
&\ge\nu_\kappa\left(\tau\rho'_{X_1}|_{E_1}+(1-\tau)\rho'_{X_2}|_{E_2}\right) \\
&\ge\tau\nu_\kappa(\rho_{X_1}'|_{E_1})+(1-\tau)\nu_\kappa(\rho_{X_2}'|_{E_2}) \\
&=\tau\nu_\kappa(\rho_{X_1}|_{E_1})+(1-\tau)\nu_\kappa(\rho_{X_2}|_{E_2}),
\end{align*}
where the first inequality follows from the Schur concavity, the second inequality follows from the concavity of the entropic function, and the last equality follows from the symmetry. 
It follows that 
\begin{align*}
\sum_{j,k}&q_j^{(1)}q_k^{(2)}\nu_\kappa(\rho_Y|_{E_1(j)E_2(k)}) \\
&\ge \tau\sum_{j}q_j^{(1)}\nu_\kappa(\rho_{X_1}|_{E_1(j)})
+(1-\tau)\sum_{k}q_k^{(2)}\nu_\kappa(\rho_{X_2}|_{E_2(k)})\\
&\ge \tau\nu_\kappa(\rho_{X_1}|_{E_1})+(1-\tau)\nu_\kappa(\rho_{X_2}|_{E_2}).
\end{align*}
This completes the proof. $\blacksquare$

\begin{table}
\caption{Summary of \textsc{EPI}s.}
\begin{ruledtabular}
\begin{tabular}{l c c  c}
{} &  $\nu_\kappa(\cdot)$  & Mixing operation  & Constant $\kappa$  \\
\hline
\textsc{cEPI} ($\star$)  &  $\surd$~\cite{B75,BL76,DCT91,R11,S59,B65,L78} & $*$ & $2/d$ \\
\textsc{qEPI} ($\star$) &   $\surd$~\cite{KS14,PMG14} & $\tau$-BS & $1/D$ \\
\textsc{qEPI}  &   $\surd$~\cite{ADO16} & $p$-\textsc{Swap} & 
$\kappa\in[0, \kappa_1]$ \\
\textsc{CqEPI} ($\star$) &  $\surd$~\cite{K15,PT17} & $\tau$-BS & $1/D$  \\
\textsc{CqEPI}  &  $\surd$~[Our proof] & $p$-\textsc{Swap} & 
$\kappa\in[0, \kappa_1]$ \\
\textsc{EPnI} ($\star$)  &  ?~\cite{GES08} & $\tau$-BS & $1/D$ [{C}]  \\
\textsc{EPnI}  &  $\surd$~\cite{ADO16} & $p$-\textsc{Swap} & 
$\kappa\in[0, \kappa_2]$ 
\end{tabular}
\end{ruledtabular}
\label{table:epi}
$\star$.~Continuous variable~(CV); $\surd$.~Hold (or proved); $*$.~Convolution; $\tau\in[0,1]$. a mixing parameter; BS.~beamsplitter; $D$.~$D$-mode; $d$.~dimensionality; 
$\kappa_1:=\tfrac{1}{(\log d)^2}$; $\kappa_2:=\tfrac{1}{d-1}$; ?.~unknown; [{C}].~conjectured.
\end{table}

\section{Discussion} \label{discussion}
In summary, we have investigated a conditional entropy power inequality for $d$-dimensional quantum systems under the assumption that ancillary environmental subsystems are separable. In the proof, we considered a post-measurement property of quantum states through a local quantum operation (especially measurement) after $p$-\textsc{Swap} on $d$-level quantum states (i.e., qudits), and applied the well-known majorization technique to the (nonincreasing order of) eigenvalues of quantum states. Our construction \textsc{CqEPI} might be useful for characterizing entanglement-assisted capacity such as for a thermal (white) noise Gaussian channel, or in quantum superdense coding.

We here discuss what is known about the entropy power inequality so far; a summary is provided in Table~\ref{table:epi}. Let us denote the entropy photon number inequality as $\textsc{EPnI}$ and the continuous variable (CV) regime by $\star$. The CV $\textsc{EPnI}$ proposed by Guha \emph{et al.} with an average photon number is an important open question in quantum Shannon theory, although recently some progress has been reported on this topic~\cite{GSG16,PTG16,PTG17}, but it is still unsolved in its original form. Furthermore, whether or not $\kappa=\tfrac{1}{D}$ on \textsc{EPnI} ($\star$) is also an important conjecture. For the $\textsc{qEPI}$ and $\textsc{CqEPI}$ on qudit versions, the entropy power inequality is still unknown for the value $\kappa=\tfrac{1}{d}$ or $\kappa>\kappa_1$. Also for the qudit $\textsc{EPnI}$ with $\kappa=\tfrac{1}{d}$ or $\kappa>\kappa_2$, the entropy power inequality is open---we do not have any strong evidence for its concave property.

Finally, we have open questions of several different kinds. For example, some dual relations on $\textsc{EPI}$ and $\textsc{qEPI}$ (and also conditional versions of $\textsc{EPI}$) in the sense of a \emph{complementary} quantum channel might be intriguing; moreover, certain inequalities of $\textsc{EPI}$s for different (or hybrid) inputs also seem to be important. It would also be interesting to study whether or not a (conditional) quantum entropy power inequality holds for quantum conditional states~\cite{LS13}, as well as for general multipartite quantum systems.

\section*{Acknowledgments}
This work was supported by the National Research Foundation of Korea (NRF) through a grant funded by the Korean government (MSIP) (Grant No. 2010-0018295) and by the KIST Institutional Program (Project No. 2E26680-16-P025). In addition, K.J. acknowledges financial support by the National Research Foundation of Korea (NRF) through a grant funded by the Korean government (Ministry of Science and ICT) (NRF-2017R1E1A1A03070510 \& NRF-2017R1A5A1015626). 
S.L. acknowledges financial support by the Basic Science Research Program through the National Research Foundation of Korea funded by the Ministry of Science, ICT \& Future Planning (NRF-2016R1A2B4014928).


\begin{thebibliography}{41}
\bibitem{BCR} 
If two input states are $D$-mode bosonic field quadratures with annihilation operators $\hat{\alpha}_1,\ldots,\hat{\alpha}_D$ and $\hat{\beta}_1,\ldots,\hat{\beta}_D$, respectively, then we obtain the $D$-mode output quadrature as $\forall j\in\{1,\ldots,D\},~\hat{\gamma}_j=\sqrt{\tau}\hat{\alpha}_j+\iota\sqrt{1-\tau}\hat{\beta}_j$.
\end{thebibliography}
\end{document}